\documentclass[aps,prl,twocolumn,superscriptaddress,showpacs,floatfix,longbibliography]{revtex4-2}
\usepackage{amsmath,amssymb,graphics,epsfig,epstopdf,color,verbatim,braket,tabularx}
\usepackage{multirow}
\usepackage{dcolumn}
\usepackage{bm}% bold math
\usepackage{amssymb}
\usepackage{supertabular}
\usepackage{tabularx}
\usepackage[colorlinks,linkcolor=blue,citecolor=blue,urlcolor=blue,bookmarks=false,breaklinks=true]{hyperref}

\begin{document}

\title{Quantized and half-quantized Anomalous Hall effect induced by in-plane magnetic field}

\author{Song Sun}
\affiliation{Beijing National Laboratory for Condensed Matter Physics, and Institute of Physics, Chinese Academy of Sciences, Beijing 100190, China}
\affiliation{University of Chinese Academy of Sciences, Beijing 100049, China}

\author{Hongming Weng}
\affiliation{Beijing National Laboratory for Condensed Matter Physics, and Institute of Physics, Chinese Academy of Sciences, Beijing 100190, China}
\affiliation{University of Chinese Academy of Sciences, Beijing 100049, China}
\affiliation{Collaborative Innovation Center of Quantum Matter, Beijing, China}

\author{Xi Dai}
\email{daix@ust.hk}
\affiliation{Department of Physics, Hong Kong University of Science and Technology, Clear Water Bay, Hong Kong}
\affiliation{Materials Department, University of California, Santa Barbara, California 93106-5050, USA}

\begin{abstract}
In this paper we propose that, quantized and nearly half-quantized intrinsic anomalous Hall effect can be induced by in-plane external magnetic field through the Zeeman coupling in non-magnetic 2D systems with sizeable spin-orbital coupling but without two-fold rotational symmetry. An analytical result is derived for 2D electron gas model with $C_{3v}$ symmetry. Based on the $\bm{k\cdot p}$ Hamiltonian derived from first principle calculations, we find that quantized and nearly half-quantized conductance can be observed in $\mathrm{Sb_2Te_3}$ thin film in the clean limit with strong in-plane magnetic field $B>20\ \mathrm{T}$ and low temperature $T<100\ \mathrm{mK}$.
\end{abstract}

\maketitle

\section{Introduction}
The Hall effect, where the transverse voltage is induced by the longitudinal current, is one of the fundamental effects for metallic systems.  It contains both the ordinary and anomalous parts of contribution, where the former is caused by the Lorentz force under an external magnetic field and the latter is due to the spin orbital effect in systems with magnetic order that breaks time reversal symmetry. The anomalous contribution to the Hall effect is called anomalous Hall effect(AHE), which can be further divided into intrinsic and extrinsic mechanisms\cite{nagaosa_rmp}. For the intrinsic mechanism, when an electric current is passing though the sample, the Bloch electrons acquire an additional anomalous velocity\cite{KarplusLuttinger1954} along the perpendicular direction, which causes the Hall current. Such an mechanism is an intrinsic property of the Bloch states in a perfect crystal without time reversal symmetry. In contrast, the extrinsic mechanism is originated from the scattering from disorder. 

Among these contributions, the intrinsic AHE has received great attention\cite{Yao2004AHCinFe,PhysRevB.74.195118,PhysRevB.76.195109,PhysRevB.75.184416} these years because it is one of the intrinsic properties for the band structures of the crystals, which can be obtained by integrating the Berry curvature over all the occupied Bloch states. The intrinsic AHE conductance in 2D can be obtained as\cite{nagaosa_rmp}
\begin{align}
  \sigma_{xy}=-\frac{e^2}{h}\sum_n\int\frac{\mathrm{d}^2\bm{k}}{2\pi}f(\epsilon_n(\bm{k}))b_n(\bm{k})\label{eq:intrinsicHall}
\end{align}
where $b_n(\bm{k})$ is the Berry curvature of the $n$th band and $f(\epsilon_n(\bm{k}))$ is the Fermi-Dirac distribution function. 
%The conductance is taken in $e^2/h$ unit through out this letter. 
For systems with time reversal symmetry, the net contribution from Eq(\ref{eq:intrinsicHall}) has to be zero, because the Berry curvature contributed by each particular Bloch state $|n,\bm{k}\rangle$ will be exactly canceled by its time reversal partner $\mathcal{T}|n, \bm{k}\rangle$. In ferro-magnetic metals, where the AHE is mostly studied, it is the magnetization $\bm{M}$ that breaks the time reversal symmetry and induces the net contribution to the Hall conductance. More over, the intrinsic AHE conductance can be possibly quantized to realise the quantum anomalous Hall effect(QAHE)\cite{ChaoxingLiu2016QAHE} when the chemical potential lies in a band gap, and the resulting band structure can be characterized by a topological index called Chern number\cite{Haldan1988PRL,Thouless1982PRL}.

Besides the ferro-magnetic metal, the AHE can also be found in non-magnetic metals or semiconductors, where the time reversal symmetry is broken not by spontaneous magnetization but the Zeeman effect caused by an external magnetic field. When the field is applied along the z-direction, the Zeeman effect is always coexisting with the Lorentz force or its quantum version, the Landau quantization, which makes it difficult to distinguish two types of the contribution. However, if we consider a 2D system and apply the field within the xy-plane, the Lorentz force or Landau quantization doesn't exist and the Hall conductance detected in such a case (if any) will be completely caused by the Zeeman effect. 

In the present paper, we will focus on such a in-plane AHE caused by the Zeeman effect. Based on the symmetry analysis, we first conclude that the in-plane AHE is commonly exist in any 2D metallic systems with sizeable spin-orbit coupling (SOC) but breaking two-fold rotational symmetry $C_{2z}$ \cite{BjYang,FangChen2015PRB}. Like the AHE in magnetic semiconductors, the in-plane AHE can also be quantized if the in-plane Zeeman effect opens a gap for the system. Such a in-plane AHE has been first discussed by professor C.X. Liu's group in magnetic topological insulator thin films\cite{ChaoxingLiu2013PRL}. Then we will propose two typical 2D metallic or semi-metallic systems which exhibit sizeable in-plane AHE, quantized in-plane AHE and more interestingly the nearly half-quantized in-plane AHE. 
The first systems is the 2D electron gas with strong SOC under point group symmetry $C_{3v}$. As we will introduced below, for such a 2D electron gas the SOC terms contains both the linear Rashba term\cite{Rashba1984, RashbaReview} and the cubic hexagonal warping term\cite{Fuliang2009PRL}, which will lead to large in-plane AHE and almost half quantized in-plane AHE. The second system discussed in detail will be the topological insulator thin films. For these materials, we first apply the first principle calculation to compute their \textit{g}-factor for the in-plane field using the method developed in our group\cite{song2015first}, which indicates a very large in-plane Zeeman effect for the bulk material already. We then calculate the sub-band structure under the in-plane magnetic field for the $\mathrm{Sb_2Te_3}$ thin film using the $\bm{k\cdot p}$ model with its parameters being extracted from the first principle calculations, based on which the Hall conductance under in-plane field can be obtained using Eq(\ref{eq:intrinsicHall}).  
Our results indicate that for such a system by adjusting two different gates on both top and bottom surfaces we can possibly reach both the in-plane quantized and nearly half-quantized AHE under strong magnetic field.

\section{The non-magnetic 2D electron gas}

Those non-magnetic 2D system can be classified into two classes by including two-fold rotational symmetry $C_{2z}$ or not. Non-zero in-plane field induced AHE can only exist in systems without $C_{2z}$ system. The reason is that, for systems with $C_{2z}$ symmetry the in-plane magnetic field breaks both the $C_{2z}$ and time reversal $T$ symmetry but preserves the combination of them, namely $C_{2z}T$ symmetry, and the Hall conductance changes sign under this $C_{2z}T$ symmetry which requires it to be zero. Hence let's focus on the 2D electron gas with $C_{3v}$ point group symmetry which excludes $C_{2z}$. The Hamiltonian of such a 2D electron gas can be written as,
\begin{align}
  H&=\frac{\hbar^2k^2}{2m}+\lambda (\bm{k}\times\bm{\sigma})\cdot \hat{z}+\frac{\delta}{2}(k_+^3+k_-^3)\sigma_z+\frac{\mu_B}{2}g\bm{\sigma}\cdot\bm{B}_\parallel,\label{eq:Hamiltonian}
\end{align}
where the first term is the kinetic term with $m$ being the effective mass, the second term is the linear Rashba coupling, the third term is the generic hexagonal warping term which breaks the $C_{2z}$ symmetry and the last term is the Zeeman coupling with $\mu_B$ being the Bohr magneton, $g$ being the in-plane effective \textit{g}-factor and $\bm{B}_\parallel$ being the in-plane magnetic field. 

Without external magnetic field, the two bands are degenerate at the $\Gamma$ point, which is protected by the time reversal symmetry. As discussed in Ref\cite{Fuliang2009PRL}, by applying an in-plane magnetic field, the degenerate point is shifted away from the $\Gamma$ point to $\bm{k}^*\equiv\frac{\mu_Bg}{2\lambda}(\hat{\bm{z}}\times \bm{B}_{\parallel})$, an energy gap $\Delta^*\equiv 2\delta{k^*}^3\sin3\theta$ ($\theta$ is the angle between $\bm{B}_{\parallel}$ and $\hat{x}$) is opened by the hexagonal warping term, except when the magnetic is exactly perpendicular to the mirror plane ($\sin3\theta=0$), and in addition the middle point of the gap is lifted up in energy $E^*\equiv\hbar^2{k^*}^2/(2m)$ by the kinetic term. Around $\bm{k}^*$, where is an avoided crossing, the Berry curvature is extremely large and contributes dominantly to the intrinsic AHE conductance. 
\begin{figure}[htbp]
  \centering\includegraphics[page=1,width=3.4in,trim={0 1.92in 4.27in 0},clip]{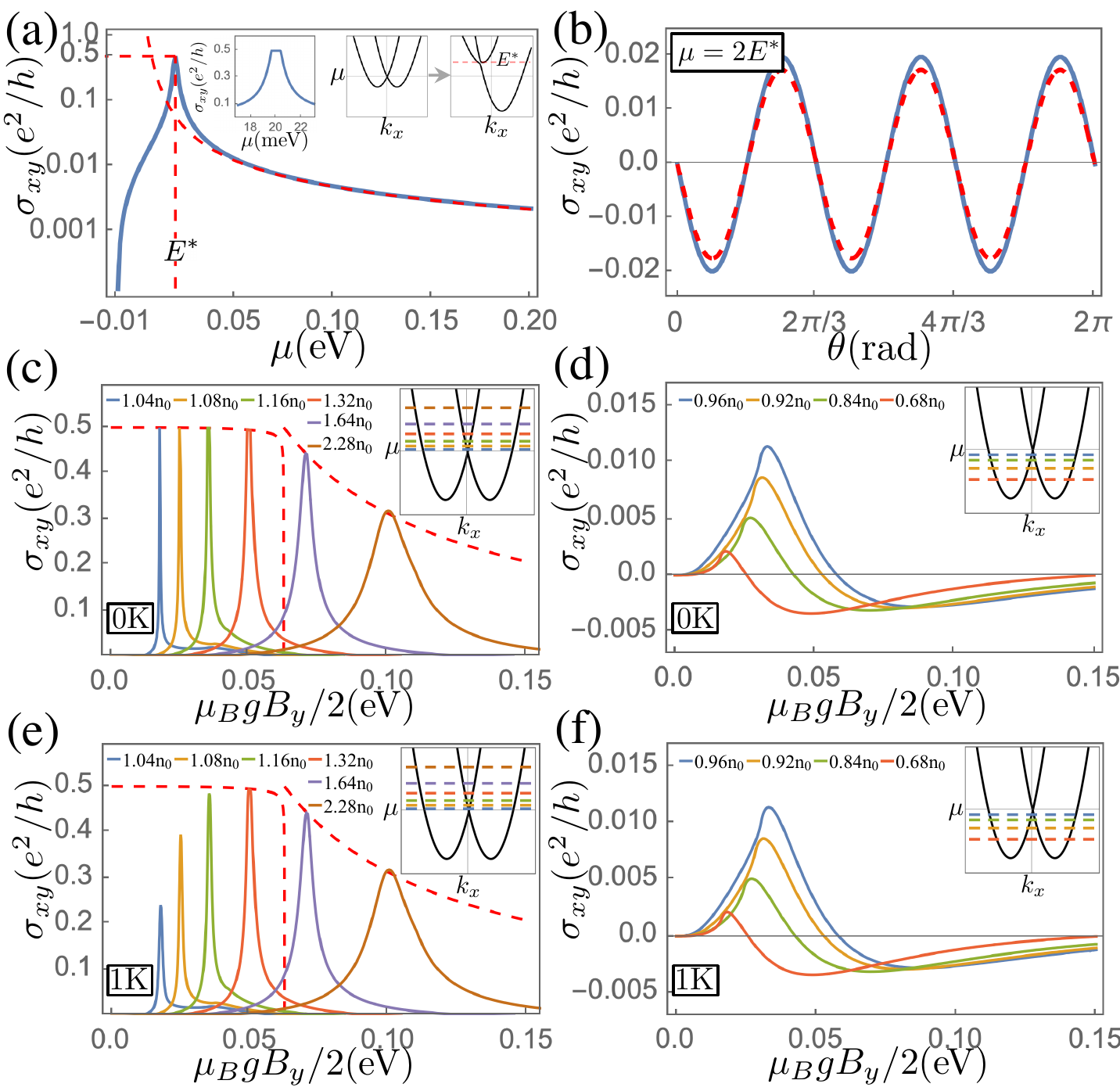}
  \caption{\label{fig:InPlaneHallAnalytic}
  The intrinsic Hall conductance of 2D electron gas with $C_{3v}$ point group symmetry. Numerical values are indicate by solid line and analytical values are indicated by red dashed line. The Hamiltonian's parameters are taken as $\hbar^2/(2m)=2$ eV$\cdot$\AA$^2$, $\lambda=0.5$ eV$\cdot$\AA\, $\delta = 0.8$ eV$\cdot$\AA$^3$, $\mu_BgB_y/2=0$ eV and $\mu_BgB_y/2=0.05$ eV. (a) and (b) shows the conductance versus the chemical potential and the angle of in-plane magnetic field respectively. In the first inset of (a), The conductance shows a nearly half-quantized plateau around $E^*$. The evolution of band structure caused by an in-plane magnetic field is shown in last two insets of (a). The zero temperature Hall conductance versus magnetic field strength with fixed electron density are shown in (c) for $n>n_0$ and (d) $n<n_0$ respectively. The insets show the position of the chemical potential in the band structure correspondingly when no magnetic field is applied. (e) and (f) shows the Hall conductance at 1 K temperature.}
\end{figure}

The hexagonal warping term\cite{Fuliang2009PRL} breaks not only the continuous rotation symmetry in z-direction down to three-fold rotation symmetry but also the vertical mirror symmetry from along all in-plane direction down to three discrete directions only. Since the vertical mirror symmetry also exclude the AHE\cite{ChaoxingLiu2013PRL}, the external magnetic field has to be applied along the direction that breaks the mirror symmetry.  

Any two bands Hamiltonian can be decomposed by Pauli matrices as $H(\bm{k}) = \epsilon(\bm{k}) + \bm{d}(\bm{k})\bm{\cdot}\bm{\sigma}$, with which the intrinsic AHE can be formulated in\cite{QiWuZhang}
\begin{align}
  \sigma_{xy}&=\int\frac{\mathrm{d}^2\bm{k}}{4\pi}\left(n_+-n_-\right)\left(\partial_{k_x} \bm{\hat{d}}\times \partial_{k_y} \bm{\hat{d}}\right)\cdot\bm{\hat{d}},
\end{align}
where $\hat{\bm{d}}(\bm{k})=\bm{d}(\bm{k})/d(\bm{k})$ is the normalized $\bm{d}$ vector and $n_{\pm}$ is the Fermi-Dirac distribution function. After some detailed calculation and approximation given in Appendix, we obtain when the chemical potential is large $\mu\gg E^*$, the leading term of the intrinsic AHE conductance at zero temperature limit is 
\begin{align}
  \sigma_{xy}\approx-\frac{\delta\mu_B^3g^3B_\parallel^3}{16\mu\lambda^3}\sin3\theta \ .
\end{align}
A result including high-order terms of $B_\parallel$ are given in the Appendix. In contrast when the chemical potential is fixed at the middle point of the gap $\mu=E^*$, the leading terms is
\begin{align}
  \sigma_{xy}\approx
  \left\{\begin{array}{ll}
    -\frac{1}{2}\frac{\sin 3 \theta}{|\sin 3 \theta|}+ \frac{\delta \mu_B^2g^2B_\parallel^2}{16\lambda^3}\frac{\sin3\theta}{\sqrt{1/\gamma^2-1}},
    & \mathrm{if} \ \gamma<1 \\
    -\frac{1}{2\gamma}\frac{\sin3\theta}{|\sin3\theta|}, & \mathrm{if}\ \gamma>1
  \end{array}\right. ,\label{eq:conductancehard}
\end{align}
where $\gamma\equiv\frac{\hbar^2\mu_BgB_\parallel}{2m\lambda^2}$ is an dimensionless quantity. When $\gamma\rightarrow 0$, which is a common case in real materials, the conductance becomes half-quantized $\sigma_{xy}\rightarrow -\frac{\sin3\theta}{2|\sin3\theta|}$. 
As shown in Fig.\ref{fig:InPlaneHallAnalytic} (a) and (b), the numerical results are in good agreement with our analytical results in these limits. We also calculate the conductance versus the magnetic field strength with fixed electron density at zero temperature shown in (c) and (d) and at 1 K temperature in (e) and (f), in which we have defined $n_0$ as the electron density when $\mu=0$ and $B=0$. Nearly half-quantized plateaux appears in (c) and (e).

The appearance of such a nearly half-quantized plateau in Hall conductance can be understood as the following. Without the cubic warping term in SOC, the continuous model described in Eq(\ref{eq:Hamiltonian}) has $C_{2z}T$ symmetry and the Zeeman coupling caused by the in-plane field shifts the band crossing point away from the $\Gamma$ point. The effect of the the hexagonal warping term then opens a small gap on the band crossing point turning it to be anti-crossing point as illustrated in the inset of Fig.\ref{fig:InPlaneHallAnalytic}(a). In the vicinity of the anti-crossing point, the band structure can be described by an asymmetric 2D Dirac model with a small mass term, which provides the half-quantized plateau feature to the Hall conductance. Of course the contribution from area away from the anti-crossing point will cause deviation from the exact half-quantized value. But our numerical calculation using the reasonable parameters introduced above strongly suggests that in the cleaning limit we can approach the half-quantized value very closely. 

Further, our results show that the Hall conductance has a nearly half-quantized plateau for $n\gtrsim n_0$ , but not for $n\lesssim n_0$. The reason is, for $n\gtrsim n_0$, when the field strength is increasing, the middle point of the gap $E^*$ is also increasing (due to the quadratic dispersion) and even go through the chemical potential if the field is strong enough. It's worth noted that, as indicated by Eq(\ref{eq:conductancehard}) and Fig.\ref{fig:InPlaneHallAnalytic} (c) and (e), when the warping term and Zeeman coupling is smaller, the conductance is really closer to half-quantized value, but the trade-off is that the gap becomes smaller, half-quantized plateau becomes thinner and measurements has to be implemented in much lower temperature. Therefore, in order to observe such half-quantized conductance, we need low temperature, strong Zeeman effect and controllable electron density.

\section{$\mathbf{Sb_2Te_3}$ Thin Film}

\begin{table*}[htbp]%tb:Bi2Se3_parameters
  \caption{\label{tb:Bi2Se3_parameters}The $\bm{k}\cdot\bm{p}$ Hamiltonian parameters for $\mathrm{Bi_2Se_3}$ family}
  %\begin{ruledtabular}
    \begin{tabular}{ccccccccc}
      \toprule
                     & $\mathrm{Bi_2Se_3}$\cite{WeiZhang_2010} & $\mathrm{Bi_2Te_3}$\cite{WeiZhang_2010} & $\mathrm{Sb_2Te_3}$\cite{WeiZhang_2010} & $\mathrm{Bi_2Se_2Te}$\cite{LinlinWang2011PRB} & $\mathrm{Bi_2Te_2Se}$\cite{LinlinWang2011PRB} & $\mathrm{BSTS}$\cite{BSTSparameter}  & $\mathrm{Sb_2Te_2Se}$\cite{Sb2Te2SeParameter} & $\mathrm{Bi_2Te_2S}$\cite{LinlinWang2011PRB} \\
    \colrule
    $A_0$(eV$\cdot$\AA)      & 3.71    & 3.45     & 3.73   & 3.47    & 3.90   & 3.23    & 3.02  & 3.87   \\
    $A_2$(eV$\cdot$\AA$^3$)  & -122.   & -172.    & -109.  & -42.6   & -294.  & -20.4   & -35.9 & -7.26  \\
    $A_3$(eV$\cdot$\AA$^3$)  & -79.2   & -60.4    & -246.  & -58.0   & -85.3  & -55.4   & -112. & -89.0  \\
    $B_0$(eV$\cdot$\AA)      & 2.48    & 0.491    & 0.967  & 1.54    & 1.70   & 1.03    & 2.25  & 1.87   \\
    $B_2$(eV$\cdot$\AA$^3$)  & -8.42   & 4.59     & 8.47   & -1.77   & -3.74  & 4.71    & -20.9 & -9.57  \\
    $B_3$(eV$\cdot$\AA$^3$)  & -241.   & -114.    & -314.  & -136.   & -293.  & -122.   & -289. & -74.4  \\
    $C_0$(eV)                & 0.0     & 0.0      & 0.0    & 0.0     & 0.0    & 0.0     & 0.0   & 0.0    \\
    $C_1$(eV$\cdot$\AA$^2$)  & 1.47    & 1.06     & -18.6  & 0.935   & -1.05  & -1.22   & -10.8 & -0.222 \\
    $C_2$(eV$\cdot$\AA$^2$)  & 22.6    & 42.9     & -17.4  & 13.7    & 66.0   & 7.46    & -4.21 & -6.95  \\
    $M_0$(eV)                & -0.218  & -0.255   & -0.200 & -0.117  & -0.345 & -0.0553 & -0.190 & -0.283\\
    $M_1$(eV$\cdot$\AA$^2$)  & 8.63    & 5.66     & 22.8   & 5.44    & 10.3   & 3.19    & 12.2  & 11.6   \\
    $M_2$(eV$\cdot$\AA$^2$)  & 45.6    & 59.4     & 58.0   & 28.0    & 91.6   & 24.7    & 49.4  & 12.3   \\
    $R_1$(eV$\cdot$\AA$^3$)  & -60.6   & -56.6    & -134.  & -21.9   & -146.  & -19.7   & -121. & 31.0   \\
    $R_2$(eV$\cdot$\AA$^3$)  & 140.2   & 125.     & 326.   & 93.5    & 221.   & 110.    & 251.  & 28.7   \\
    $R_3$(eV$\cdot$\AA$^3$)  & 174.2   & 264.     & 490.   & 150.    & 356.   & 143.    & 229.  & 51.5   \\
    $g_{1z}$                 & -23.6   & -49.4    & -17.5  & -16.9   & -74.7  & -11.6   & -10.4 & 3.36   \\
    $g_{1p}$                 & -5.89   & -4.99    & -4.60  & -4.52   & -5.43  & -3.69   & -4.95 & -4.89  \\
    $g_{2z}$                 & 5.13    & 9.31     & 15.9   & 8.58    & 8.48   & 7.98    & -0.219 & 8.17  \\
    $g_{2p}$                 & -6.77   & -5.49    & -19.9  & -5.63   & -7.93  & -5.15   & -16.0 & -7.30  \\
    %Lattice Constants & \cite{WeiZhang_2010} & \cite{WeiZhang_2010} & \cite{WeiZhang_2010} & \cite{LinlinWang2011PRB} & \cite{LinlinWang2011PRB} & \cite{ZhiRen2010PRB} & \cite{Sb2Te2SeParameter} & \cite{LinlinWang2011PRB} \\
    \botrule
    \end{tabular}
  %\end{ruledtabular}
\end{table*}

\begin{figure}[htbp]
  \centering\includegraphics[page=2,width=3.4in,trim={0 2.85in 3.25in 0},clip]{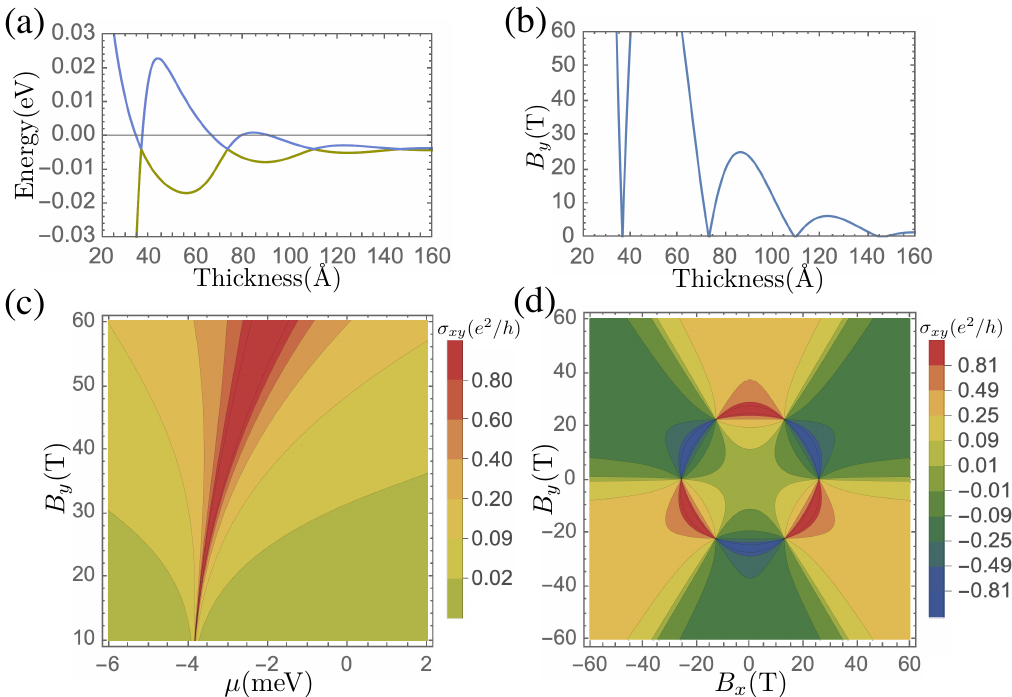}
  \caption{\label{fig:Sb2Te3line} 
  Sub-bands structure and in-plane magnetic field induced QAHE in $\mathrm{Sb_2Te_3}$ thin film: 
  (a) The energy levels of sub-bands at the $\Gamma$ point versus the thickness of the thin film. (b) the minimum strength of in-plane magnetic field to realize the QAHE versus the thickness of thin film. The magnetic field is fixed in $\hat{y}$ direction. 
  The in-plane magnetic field induced intrinsic AHE conductance of Sb$_2$Te$_3$ thin film with thickness being 110 \AA are shown in (c) and (d). (c) the intrinsic AHE conductance versus chemical potential and $\hat{y}$ directional in-plane magnetic field strength. (d) the intrinsic AHE conductance versus in-plane magnetic field strength in both $\hat{x}$ and $\hat{y}$ direction. The chemical potential is being set as -3.5 meV}
\end{figure}

The $\mathrm{Bi_2Se_3}$ family of compounds including $\mathrm{Bi_2Se_3}$, $\mathrm{Bi_2Te_3}$, $\mathrm{Sb_2Te_3}$\cite{HaijunZhang2009NP,Xia2009,WeiZhang_2010,Analytis2010NP,Xia2009}, $\mathrm{Bi_2Se_2Te}$, $\mathrm{Bi_2Te_2Se}$, $\mathrm{Bi_2Te_2S}$\cite{ZhiRen2010PRB,LinlinWang2011PRB}, $\mathrm{BiSbTeSe_2}$\cite{Arakane2012NC, Xia2013PRB, Segawa2012PRB, YongChen2014NP} and $\mathrm{Sb_2Te_2Se}$ \cite{Sb2Te2SeParameter,Lee2016} are all three dimensional strong TI with large bulk band gap predicted by first principle calculations and confirmed by experiments. 
The crystal structure of these materials are rhombohedral with layered structure belonging to $R\bar{3}m$ space group and $D_{3d}$ point group. As discussed in \cite{ModelHamilt,HaijunZhang2009NP}, with symmetry principles and analysis of the atomic orbitals, the model $\bm{k}\cdot\bm{p}$ Hamiltonian up to third order of $\bm{k}$ around $\Gamma$ point can be constructed as following,
\begin{align}    
  H_0=&\epsilon+\mathcal{M}\sigma_z+\mathcal{B}k_z\sigma_y+\mathcal{A}(k_ys_x\sigma_x-k_xs_y\sigma_x) \nonumber\\    
  &+\mathcal{R}_1s_z\sigma_x+\mathcal{R}_2\sigma_y
  %&+R_3k_z\left(\frac{k_+^2+k_-^2}{2}s_x\sigma_x-\frac{k_+^2-k_-^2}{2i}s_y\sigma_x\right)\\
  +\mathcal{R}_{3x}s_x\sigma_x-\mathcal{R}_{3y}s_y\sigma_x ,
\end{align}
where $\epsilon({\bm{k}})=C_{0}+C_1k_z^2+C_2k_\parallel^2$, $\mathcal{M}(\bm{k})=M_0+M_1k_z^2+M_2k_\parallel^2$, $\mathcal{A}(\bm{k})=A_0+A_2k_\parallel^2+A_3k_z^2$, $\mathcal{B}(\bm{k})=B_0+B_2k_z^2+B_3k_{\parallel}^2$, $\mathcal{R}_1(\bm{k})=R_1\frac{k_+^3+k_-^3}{2}$, $\mathcal{R}_2(\bm{k})=R_2\frac{k_+^3-k_-^3}{2i}$, $\mathcal{R}_{3x}=R_3\frac{(k_+^2+k_-^2)k_z}{2}$, $\mathcal{R}_{3y}=R_3\frac{(k_+^2-k_-^2)k_z}{2i}$, $k_{\parallel}^2=k_x^2+k_y^2$. And the Zeeman coupling is
\begin{align}
  H_z=&\frac{\mu_B}{2}\left[\tilde{g}_{1z}B_zs_z+\tilde{g}_{1p}\left(B_xs_x+B_ys_y\right)\right. \nonumber \\
  &+\left.\tilde{g}_{2z}B_z\sigma_zs_z+\tilde{g}_{2p}\sigma_z\left(B_xs_x+B_ys_y\right)\right] ,
  %H_z=\frac{\mu_B}{2}&\left[(\tilde{g}_{1z}+\tilde{g}_{2z}\sigma_z)B_zs_z+(\tilde{g}_{1p}+\tilde{g}_{2p}\sigma_z)\left(B_xs_x+B_ys_y\right)\right]
\end{align}
where $\tilde{g}_{1z}=(g_{1z}+g_{2z})/2$, $\tilde{g}_{1p}=(g_{1p}+g_{2p})/2$, $\tilde{g}_{2z}=(g_{1z}-g_{2z})/2$ and $\tilde{g}_{2p}=(g_{1p}-g_{2p})/2$.
%The irreducible representation of k and matrix of the D3d group are given in Appendix.
In this work we have recalculated all the parameters of this model together with the \textit{g}-factors using the first principle method developed previously by our group\cite{song2015first} for all of the $\mathrm{Bi_2Se_3}$ family of compounds listed above, which are summarized in Table.\ref{tb:Bi2Se3_parameters}. Among these materials, the $\mathrm{Bi_2Te_2Se}$ has the largest out of plane \textit{g}-factor being 74.7 and the $\mathrm{Sb_2Te_3}$ has the largest in-plane \textit{g}-factor being 19.9. Due to the large in-plane \textit{g}-factor and hexagonal warping term in $\mathrm{Sb_2Te_3}$, we take it as an example in the following to show that the intrinsic AHE conductance, including quantized and nearly half-quantized AHE, can all be realized in these materials.

\begin{figure*}[htbp]
  \centering\includegraphics[page=3,width=6.in,trim={0.in 1.70in 1.55in 0},clip]{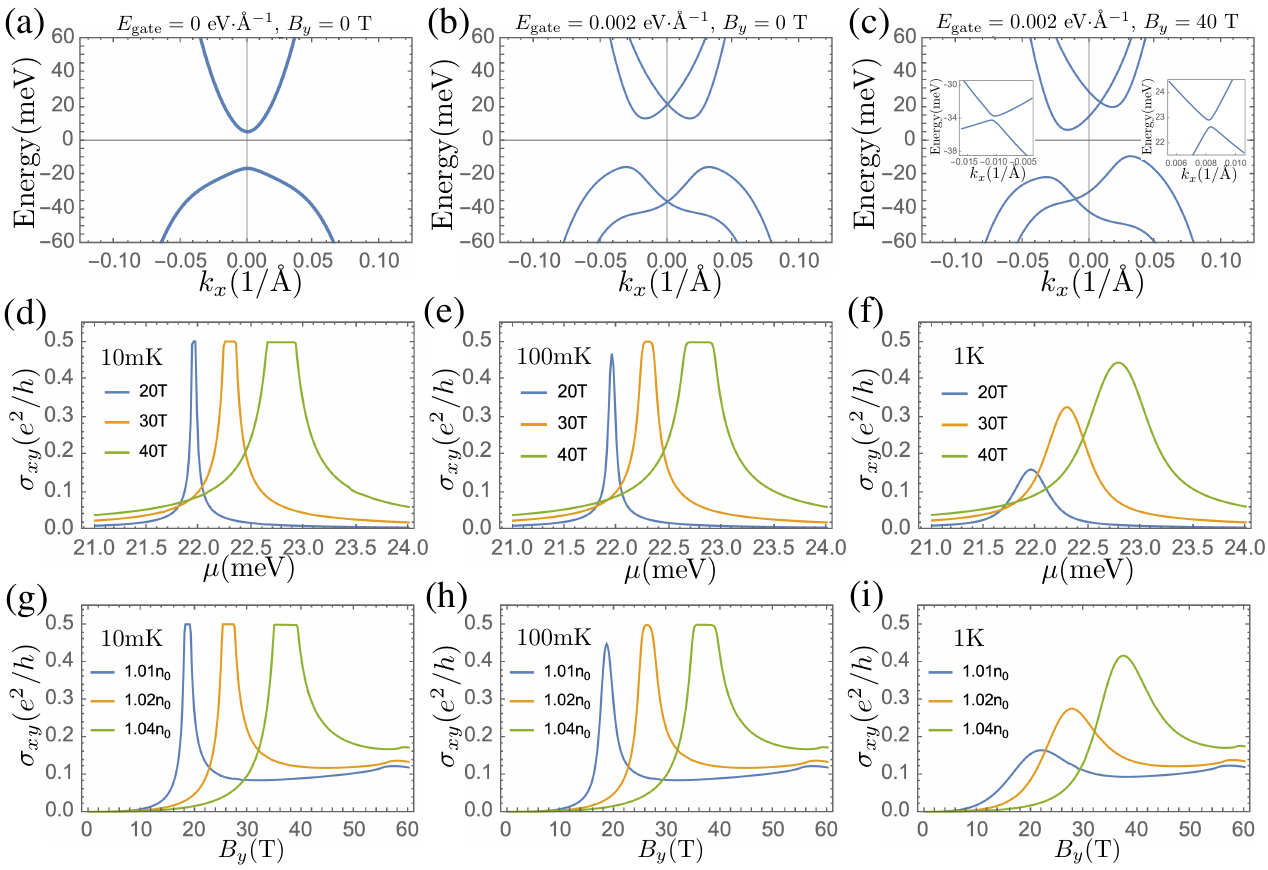}
  \caption{\label{fig:Sb2Te3electric}
  The band structure and in-plane intrinsic AHE conductance for 60\AA\ thick $\mathrm{Sb_2Te_3}$ thin film with $E_{\mathrm{gate}}=0.002$ eV$\cdot$\AA$^{-1}$. (a)$\sim$(c) show the evolution of band structure by introducing gate tuning and in-plane magnetic field. Insets of (c) shows a gap is opened by the warping term. (d)$\sim$(f) are in-plane AHE conductance versus chemical potential with different field strength and temperature respectively (g)$\sim$(i) are in-plane AHE conductance versus field strength with different electron density and temperature respectively.}
\end{figure*}

By treating the thin film as an quantum well with infinite potential barrier, the sub-band Hamiltonian can be derived by envelope function approach\cite{WINKLER20122096,Burt_1992} from 4 bands bulk Hamiltonian given above, where the basis functions are taken as $\psi_n=\sqrt{\frac{2}{L}}\sin\left({\frac{n\pi z}{L}+\frac{n\pi}{2}}\right)$ with $L$ being the thickness of the thin film. 

As shown in Fig.\ref{fig:Sb2Te3line} (a), with the reduction of the film thickness, as a result of quantum confinement effect the two lowest energy sub-bands with opposite parity cross each other repeatedly\cite{Oscillatory,HaizhouLu2010,Linder2009,Shan_2010}. This oscillatory behavior allows us to tune the sub-band gap by thickness.
When the magnetic field is applied in $\hat{y}$ direction, which is the bisector between the two neighboring mirror planes, we find that the required in-plane field strength to realize QAHE also oscillates with the thickness and reaches zeros at those transition points between topological and non-topological phases as shown in Fig.\ref{fig:Sb2Te3line} (b), and we can find that for film thickness around 4 nm, 8 nm and 11 nm we have the best chances to realize QAHE by in-plane Zeeman effect under the feasible field strength.
We also plot the intrinsic AHE conductance at zero temperature as a function of magnetic field strength and chemical potential in Fig.\ref{fig:Sb2Te3line} (c), which shows that with strong enough in-plane magnetic field and fine-tuning of chemical potential the quantized in-plane Hall effect can be realized in the thin films proposed in the present study. The mechanism behind the Chern insulator phases here can still be ascribed to the band inversions at the $\Gamma$ point. The Zeeman splitting induced by the strong in-plane magnetic field will overcome the gap and invert only a single pair of bands with opposite parity at $\Gamma$ leading to nonzero Chern number as illustrated in more detail in reference\cite{RuiYu-2010-science,ChaoxingLiu2013PRL}.
We also calculated the Hall conductance versus the in-plane magnetic field strength with fixing chemical potential in Fig.\ref{fig:Sb2Te3line} (d) which shows that we can modify the sign of intrinsic AHE by varying the direction of in-plane magnetic field.  

For such thin film systems, we can further introduce a vertical electric field by setting different gate voltage on top and bottom surfaces to break the inversion symmetry, which reduces the point group from $D_{3d}$ to $C_{3v}$. As shown in Fig.\ref{fig:Sb2Te3electric} (b) and (c), the sub-band structure of two conduction bands resembles the 2D electron gas with $C_{3v}$ symmetry given above, which implies that the half-quantized in-plane AHE can also be realized here if the chemical potential is tuned to the right position. The electric field is taken into account by adding a diagonal $zE_{\mathrm{gate}}$ term in the quantum well Hamiltonian. We have calculated the in-plane AHE conductance using the above method and the results are shown in (d)$\sim$(i), from which we conclude that with strong in-plane magnetic field $B>20\ \mathrm{T}$ and low temperature $T<100\ \mathrm{mK}$ the nearly half-quantized conductance can be observed in such a thin film.

\section{Summary}
In this letter, we propose that the the in-plane Zeeman effect induced AHE commonly exists in 2D materials systems with strong SOC but without $C_{2z}$ symmetry. By tuning the field strength and chemical potential, not only quantized but also nearly half-quantized AHE can be realized in these 2D systems, which is completely new in condensed matter physics. Two material examples has been chosen to carry out numerical calculations. The first one is the 2D electron gas with Rashba type SOC and three-fold rotational symmetry. Large in-plane AHE can be realized in such a generic 2D metallic systems with the highest in-plane Hall conductance being close to the half quantized value, which can be approached by tuning both the field strength and the carrier density. The second material example is the $\mathrm{Bi_2Se_3}$ family thin films, among which we find that $\mathrm{Sb_2Te_3}$ has the largest in-plane Zeeman effect. Finally, we numerically studied the in-plane Hall effect in $\mathrm{Sb_2Te_3}$ thin film in detail and show that both the quantized and nearly half-quantized AHE conductance can be reached.

\section{Acknowledgments}
X.D. acknowledges financial support from the Hong Kong Research Grants Council (Project No. GRF16300918 and No. 16309020).
H.W. and S.S. acknowledge the supports from the National Natural Science Foundation (Grant No. 11925408, 11921004 and 12188101), the Ministry of Science and Technology of China (Grant No. 2018YFA0305700), the Chinese Academy of Sciences (Grant No. XDB33000000), the K. C. Wong Education Foundation (GJTD-2018-01), and the Informatization Plan of Chinese Academy of Sciences, (Grant No. CAS-WX2021SF-0102).
%
%$\sigma_{xy}(e^2/h)$ \ $\mu$\ $E^*$\ $\mu$(meV)
%
%$k_x$ $\mu=2E^*$\ $\theta$(rad)\ $\mu_BgB_y/2$(eV)\ $0$K\ $1$K\ 
%
%$\mu$(eV) 
%
%$B_x$(T)\ $B_y$(T)\ Thickness(\AA)\ Energy(eV)
%
%Energy(meV)\ $k_x$(eV$\cdot$\AA)\ 
%
%$E_\mathrm{gate}=0$ eV$\cdot$\AA$^{-1}$, $B_y=0$ T 
%
%$E_\mathrm{gate}=0.002$ eV$\cdot$\AA$^{-1}$, $B_y=0$ T
%
%$E_\mathrm{gate}=0.002$ eV$\cdot$\AA$^{-1}$, $B_y=40$ T
%
%$k_x$(1/\AA)\ 10mK 100mK 1K
%
%$k_x$ $k_y$
%
%$\mu\gg E^{*}$ $\mu=E^{*}$, $\gamma<1$ $\mu=E^{*}$, $\gamma>1$

\bibliography{Manuscript}
\clearpage
\appendix
\begin{widetext}
\section{In-plane Hall effect in quasi-2D metals with $\mathrm{C_{3v}}$ symmetry}
%\subsubsection{Proof for the equivalence of Berry curvature expression}
For general two-band 2D systems, the $\bm{k\cdot p}$ Hamiltonian can be formulated by $H(\bm{k}) = \epsilon(\bm{k}) + \bm{d}(\bm{k})\bm{\cdot}\bm{\sigma}$, where $\bm{d}=(d_x,d_y,d_z)$, $\bm{\sigma}=(\sigma_{x},\sigma_{y},\sigma_{z})$ and $\bm{k}=(k_x,k_y)$. 
With Kubo formula, the Hall conductance is derived as \cite{QiWuZhang}
\begin{align}
  %\sigma_{xy}=\frac{1}{4\pi}\frac{e^2}{h}\int_{FBZ}\left(n_+(\bm{k})-n_-(\bm{k})\right)\left(\frac{\partial \bm{\hat{d}}}{\partial k_x}\times \frac{\partial \bm{\hat{d}}}{\partial k_y}\right)\cdot\bm{\hat{d}} \ \mathrm{d}^2 \bm{k} \\
  \sigma_{xy}=\frac{1}{4\pi}\int\left(n_+(\bm{k})-n_-(\bm{k})\right)\left(\partial_{k_x} \bm{\hat{d}}\times \partial_{k_y} \bm{\hat{d}}\right)\cdot\bm{\hat{d}} \ \mathrm{d}^2 \bm{k} ,
\end{align}
where $\hat{\bm{d}}(\bm{k})$ is the normalized $\bm{d}(\bm{k})$ vector. Geometrically, $(\partial_{k_x} \bm{\hat{d}}\times \partial_{k_y} \bm{\hat{d}})\cdot\bm{\hat{d}} \ \mathrm{d}^2 \bm{k}$ describes the differential solid angle in $\bm{d}$ space, therefore the normalization factor can be taken out of the derivative operators. Therefore, by utilizing $\bm{d}\times \hat{\bm{d}}=0$, it's straightforward to prove that
\begin{align}
  \left(\partial_{k_x} \bm{\hat{d}}\times \partial_{k_y} \bm{\hat{d}}\right)\cdot\bm{\hat{d}} = \left(\partial_{k_x} \bm{d} \times \partial_{k_y} \bm{d} \right)\cdot \bm{d}/d^3 .
\end{align}
In the polar coordinate with $k=|\bm{k}|$ and $\varphi=\arctan(k_y/k_x)$, the integral becomes
\begin{align}
  \sigma_{xy}=\frac{1}{4\pi}\int\left(n_+(\bm{k})-n_-(\bm{k})\right)\left(\partial_k \bm{d}\times \partial_\varphi \bm{d}\right)\cdot\bm{d}/d^{3} \ \mathrm{d}k \mathrm{d}\varphi . \label{eq:derivatives}
\end{align}

%\begin{align}
%    b&=\frac{1}{2}\left(\frac{\partial \bm{\hat{d}}}{\partial k_x}\times \frac{\partial \bm{\hat{d}}}{\partial k_y}\right)\cdot\bm{\hat{d}} \\
%    &=\frac{1}{2}\left(\frac{\partial (\bm{d}/d)}{\partial k_y}\times\bm{\hat{d}}\right)\cdot\frac{\partial \bm{\hat{d}}}{\partial k_x}\\
%    &=\frac{1}{2}\left[\left(\frac{1}{d}\frac{\partial \bm{d}}{\partial k_y}+\frac{\partial 1/d}{\partial k_y}\bm{d}\right)\times\bm{\hat{d}}\right]\cdot\frac{\partial \bm{\hat{d}}}{\partial k_x}\\
%    &=\frac{1}{2d}\left(\frac{\partial \bm{d}}{\partial k_y}\times\bm{\hat{d}}\right)\cdot\frac{\partial \bm{\hat{d}}}{\partial k_x} \\
%    &=\frac{1}{2d}\left(\bm{\hat{d}}\times\frac{\partial \bm{\hat{d}}}{\partial k_x}\right)\cdot \frac{\partial \bm{d}}{\partial k_y}\\
%    &=\frac{1}{2d}\left(\bm{\hat{d}}\times\frac{\partial (\bm{d}/d)}{\partial k_x}\right)\cdot \frac{\partial \bm{d}}{\partial k_y} \\
%    &=\frac{1}{2d^3}\left( \frac{\partial \bm{d}}{\partial k_x} \times \frac{\partial \bm{d}}{\partial k_y} \right)\cdot \bm{d}
%\end{align}

Now we focus on deriving $\bm{d}(\bm{k})$ for the Hamiltonian given in the main text. Let's first rewrite the Hamiltonian in matrix form 
\begin{align}
  H&=\frac{\hbar^2k^2}{2m}+\lambda (\bm{k}\times\bm{\sigma})\cdot \bm{e}_z+\frac{\delta}{2}(k_+^3+k_-^3)\sigma_z+\frac{\mu_B}{2}g\bm{\sigma}\cdot\bm{B} \\
  &=\frac{\hbar^2k^2}{2m}+
   \left(\begin{array}{cc}
       \frac{\delta}{2}(k_+^3+k_-^3) & -i\lambda k_- +\mu_BgB_-/2\\
       i\lambda k_+ + \mu_BgB_+/2& -\frac{\delta}{2}(k_+^3+k_-^3)
   \end{array}\right).
\end{align}
Because the Rashba term is shifted by the Zeeman coupling and the warping term is always a perturbation, we can hide complexity in the warping term by taking a shift over $\bm{k}$ with $k_x\rightarrow k_x-\mu_BgB_y/(2\lambda)$ and  $k_y\rightarrow k_y+\mu_BgB_x/(2\lambda)$. And transforming it into polar coordinate with $\varphi=\arctan\frac{k_y}{k_x}$ and $\theta=\arctan\frac{B_y}{B_x}$, we obtain
\begin{align}
  \epsilon(\bm{k})&=\frac{\hbar^2}{2m}\left(k^2+\frac{\mu_B^2g^2B^2}{4\lambda^2} + k\frac{\mu_BgB}{\lambda}\sin(\varphi-\theta)\right),\\
  d_x(\bm{k})&=-\lambda k \sin\varphi ,\\
  d_y(\bm{k})&=\lambda k \cos\varphi ,\\
  d_z(\bm{k})&=\delta \left(k^3\cos3\varphi-k^2\frac{3\mu_BgB}{2\lambda}\sin(2\varphi+\theta)-k\frac{3\mu_B^2g^2B^2}{4\lambda^2}\cos(\varphi+2\theta) + \frac{\mu_B^3g^3B^3}{8\lambda^3}\sin3\theta\right) ,
\end{align}
By directly calculating the mixed product, we obtain 
\begin{align}
  (\partial_k \bm{d} \times \partial_\varphi \bm{d}) \cdot \bm{d} &= \delta\left(k \frac{\mu_B^3g^3B^3}{8\lambda}\sin 3\theta + k^3\frac{3\lambda \mu_BgB}{2}\sin(\theta+2\varphi) - 2k^4 \lambda^2 \cos 3\varphi\right).
\end{align}
The mixed product is consist of three terms named $k^1$, $k^3$ and $k^4$ term. In order to complete the integrate in Eq(\ref{eq:derivatives}), we have to make some approximations to $d(\bm{k})$. It is worth noting that the variable of integration $k$ goes from $0$ to $k_F$(the Fermi surface) and the integrand is $(\partial_k \bm{d} \times \partial_\varphi \bm{d})/d^3$. Hence for $k^1$ term, we take $d(\bm{k})\approx \sqrt{\lambda^2k^2+\frac{\delta^2\mu_B^6g^6B^6\sin^23\theta}{64\lambda^6}}$ when $k\lesssim \frac{\delta\mu_B^3g^3B^3|\sin3\theta|}{8\lambda^4}$ to prevent the divergence and approximate it further as $d(\bm{k})\approx \lambda k$ when $k\gg\frac{\delta\mu_B^3g^3B^3|\sin3\theta|}{8\lambda^4}$. And we can always take $d(\bm{k})\approx \lambda k$ for $k^3$ and $k^4$ term. After these approximations we obtain the integrand as
\begin{align}
  (\partial_k \bm{d} \times \partial_\varphi \bm{d}) \cdot \bm{d} /d^3 \approx  k\frac{\frac{\delta\mu_B^3g^3B^3\sin 3\theta}{8\lambda^3}}{(\lambda^2 k^2 + \frac{\delta^2\mu_B^6g^6B^6\sin^23\theta}{64\lambda^6})^{\frac{3}{2}}} 
  + \frac{3\delta\mu_BgB}{2\lambda^2}\sin(\theta+2\varphi) - k \frac{2\delta}{\lambda} \cos 3\varphi \ . \label{eq:curvatureapp}
\end{align}
Doing the integrate over $k$ yields
\begin{align}
  \sigma_{xy}&=\frac{1}{4\pi}\int_{0}^{2\pi} \int_{k_{F1}}^{k_{F2}} (\partial_k \bm{d} \times \partial_\varphi \bm{d}) \cdot \bm{d}/d^3 \ \mathrm{d}\bm{k} \mathrm{d}\bm{\varphi} \\
   &\approx \frac{1}{4\pi}\int_{0}^{2\pi} \left[-\frac{\frac{\delta\mu_B^3g^3B^3\sin3\theta}{8\lambda^3}}{\sqrt{\lambda^2 k^2 + \frac{\delta^2\mu_B^6g^6B^6\sin^23\theta}{64\lambda^6}}} + k\frac{3\delta\mu_BgB}{2\lambda^2}\sin(\theta+2\varphi) - k^2 \frac{\delta}{\lambda} \cos 3\varphi\right]_{k_{F1}}^{k_{F2}}\mathrm{d}\bm{\varphi} \label{eq:integratek}
\end{align}
where $k_{F1}$ and $k_{F2}$ representing Fermi surfaces are functions of $\varphi$ in general. 

%Now we have to discuss the calculation  for $\mu>>\frac{\hbar^2\mathfrak{B}^2}{2m^*\lambda^2}$ and $\mu=\frac{\hbar^2\mathfrak{B}^2}{2m^*\lambda^2}$ respectively. 
\begin{figure*}[htbp]
  \centering\includegraphics[page=4,width=5.8in,trim={0 4.80in 1.6in 0},clip]{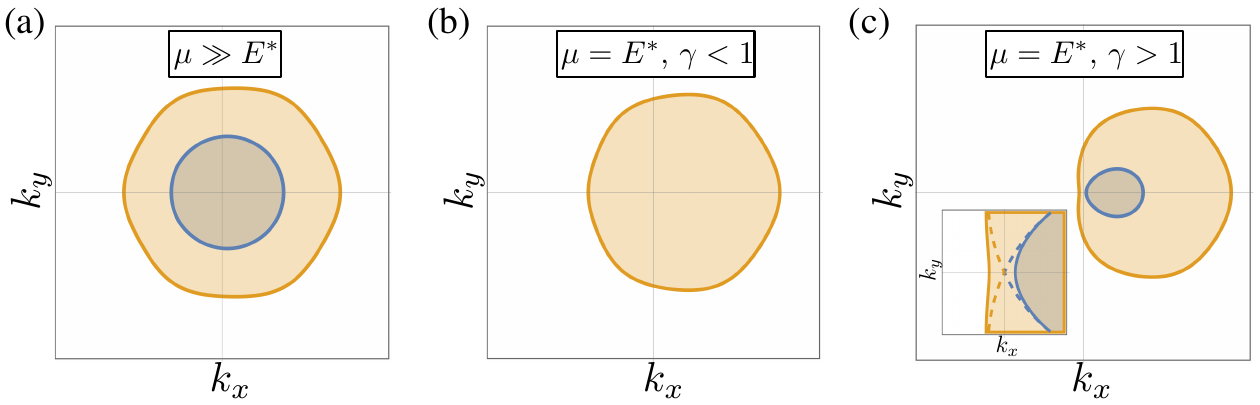}
  \caption{\label{fig:fs}
  Schematics diagrams to show the shape of Fermi surface under three different conditions. Inset of (c) is a enlarged view of the center region to show the region needs to be integrated additionally.}
\end{figure*}
Let's firstly focus on $\mu \gg E^*\equiv\hbar^2{k^*}^2/(2m)$. The shape of Fermi surface is shown in Fig.\ref{fig:fs}(a). To a first-order approximation of $\delta$, Fermi surfaces are approximated by
\begin{align}
  k_{F1}&\approx-\frac{\mu_BgB}{2\lambda}\sin(\varphi-\theta)+\frac{m\lambda}{\hbar^2}+\sqrt{\left(\frac{\mu_BgB}{2\lambda}\sin(\varphi-\theta)-\frac{m\lambda}{\hbar^2}\right)^2-\left(\frac{\mu_B^2g^2B^2}{4\lambda^2}-\frac{2m\mu}{\hbar^2}\right)} \ , \label{eq:kF1}\\
  k_{F2}&\approx-\frac{\mu_BgB}{2\lambda}\sin(\varphi-\theta)-\frac{m\lambda}{\hbar^2}+\sqrt{\left(\frac{\mu_BgB}{2\lambda}\sin(\varphi-\theta)+\frac{m\lambda}{\hbar^2}\right)^2-\left(\frac{\mu_B^2g^2B^2}{4\lambda^2}-\frac{2m\mu}{\hbar^2}\right)} \ . \label{eq:kF2}
\end{align}
We can expand $k_F$ as polynomials of $B$. Particularly to the accuracy of zeroth-order of $B$, we have 
\begin{align}
  k_{F1}&\approx\frac{\sqrt{2 \mu \hbar^2/m+\lambda ^2}+\lambda }{\hbar ^2/m},\\
  k_{F2}&\approx\frac{\sqrt{2 \mu \hbar^2/m+\lambda ^2}-\lambda }{\hbar ^2/m}.
\end{align} 
Plugging them into Eq(\ref{eq:integratek}), it straightforward to get the leading term of the Hall conductance as 
\begin{align}
  \sigma_{xy}\approx-\frac{\delta\mu_B^3g^3B^3}{16\mu\lambda^3}\sin3\theta, \ \mathrm{if}\ \mu\gg E^*.
\end{align}
Further, if we retain more high order terms of $B$ in the series expansion of $k_F$ in Eq(\ref{eq:kF1}) and Eq(\ref{eq:kF2}), we will get a better approximation as 
\begin{align}
  \sigma_{xy}\approx -\frac{ \delta \mu_B^3g^3B^3\sin 3\theta}{16\mu\lambda^3}\left(1 + \frac{\mu_B^2g^2B^2\hbar^2}{8\mu\lambda^2m} + \frac{\mu_B^4g^4B^4\hbar^4}{64\mu^2\lambda^4m^2}\right), \ \mathrm{if}\ \mu\gg E^*.
\end{align}

We now discuss the calculation for $\mu=E^*$. In this case, when $\gamma\equiv\frac{\hbar^2\mu_BgB}{2m\lambda^2}<1$, only the lower band has Fermi surface and the higher band is fully unoccupied as shown in Fig.\ref{fig:fs} (b) The Fermi surface is approximated by
\begin{align}
  k_{F1}&\approx 2\left(\frac{m\lambda}{\hbar^2}-\frac{\mu_BgB}{2\lambda}\sin(\varphi-\theta)\right) , \\%\ \mathrm{with} \ 0<\varphi\leq 2\pi\\
  k_{F2}&\approx 0 \ .
\end{align}
Plugging them into  Eq(\ref{eq:integratek}) we derive
\begin{align}
  \sigma_{xy}&\approx \frac{1}{4\pi}\int_{0}^{2\pi}\left(\frac{\delta\mu_B^3g^3B^3\sin3\theta}{8\lambda^4k_{F1}}-\frac{\sin3\theta}{|\sin3\theta|}\right)\mathrm{d}\varphi\\
  &=-\frac{1}{2}\frac{\sin 3\theta}{|\sin 3\theta|}+\frac{1}{8\pi}\frac{\delta\mu_B^2g^2B^2\gamma\sin3\theta}{4\lambda^3}\int_{0}^{2\pi}\frac{\mathrm{d}\varphi}{1-\gamma\sin(\varphi-\theta)}
\end{align}
Here, for $k_{F1}$ we have used the further approximation $\sqrt{\lambda^2k^2+\frac{\delta^2\mu_B^6g^6B^6\sin^23\theta}{64\lambda^6}}\approx \lambda k$. 
By transforming it to complex contour integral, %With transformation $\sin\theta=(z-z^{-1})$ and $\mathrm{d}\varphi=\mathrm{d}z/iz$, 
the integral can be derived with the residual theorem. Finally we get
\begin{align}
  \sigma_{xy}\approx-\frac{1}{2}\frac{\sin 3 \theta}{|\sin 3 \theta|}+\frac{\delta\mu_B^2g^2B^2}{16\lambda^3}\frac{\sin3\theta}{\sqrt{1/\gamma^2-1}}, \ \mathrm{if}\ \mu=E^* \ \mathrm{and} \ \gamma<1 \ .
\end{align}

In contrast when $\gamma>1$ both bands have occupied region as shown in Fig.\ref{fig:fs}(c) The Fermi surface is approximated by
\begin{align}
  k_{F1}\approx&\left\{\begin{array}{ll} \frac{2m\lambda}{\hbar^2}-\frac{\mu_BgB}{\lambda}\sin(\varphi-\theta), & \quad \mathrm{if} \ \varphi-\theta \in \left(-\pi-\alpha,\alpha\right) \\
    0, &\quad \mathrm{otherwise}
  \end{array}\right. , \label{eq:FShard1}\\
  k_{F2}\approx&\left\{\begin{array}{ll} -\frac{2m\lambda}{\hbar^2}-\frac{\mu_BgB}{\lambda}\sin(\varphi-\theta), & \mathrm{if} \ \varphi-\theta \in \left(-\pi+\alpha,-\alpha\right) \\
    0, & \mathrm{otherwise}
  \end{array}\right. , \label{eq:FShard2}
\end{align}
where $\alpha\equiv\arcsin(\frac{1}{\gamma})$. Taking them into Eq(\ref{eq:integratek}) we obtain
\begin{align}
  \sigma_{xy}^0&\approx -\frac{1}{4\pi}\left(\int_{-\pi+\alpha}^{-\alpha}\left[\frac{\frac{\delta\mu_B^3g^3B^3\sin3\theta}{8\lambda^3}}{\sqrt{\lambda^2 k^2 + \frac{\delta^2\mu_B^6g^6B^6\sin^23\theta}{64\lambda^6}}}\right]_{k_{F1}}^{k_{F2}}\mathrm{d}\varphi +\left(\int_{-\pi-\alpha}^{-\pi+\alpha}+\int_{-\alpha}^{\alpha}\right) \left[\frac{\frac{\delta\mu_B^3g^3B^3\sin3\theta}{8\lambda^3}}{\sqrt{\lambda^2 k^2 + \frac{\delta^2\mu_B^6g^6B^6\sin^23\theta}{64\lambda^6}}}\right]_{k_{F1}}^{0}\mathrm{d}\varphi\right) \\
  &\approx -\frac{1}{4\pi}\left(\int_{-\pi+\alpha}^{-\alpha}  \frac{\delta\mu_B^3g^3B^3\sin 3\theta}{8\lambda^3}\left(\frac{1}{k_{F2}}-\frac{1}{k_{F1}}\right)\mathrm{d}\varphi
   +\left(\int_{-\pi-\alpha}^{-\pi+\alpha}+\int_{-\alpha}^{\alpha}\right) \left(\frac{\sin 3\theta}{|\sin 3\theta|}-\frac{\delta \mu_B^3g^3B^3\sin 3\theta}{8\lambda^3}\frac{1}{k_{F1}}\right)\mathrm{d}\varphi\right) \\
  &=-\frac{\alpha}{\pi}\frac{\sin3\theta}{|\sin3\theta|}-\frac{\delta}{4\pi}\frac{\mu_B^3g^3B^3\sin3\theta}{8\lambda^4}\left(\int_{-\pi+\alpha}^{-\alpha}\frac{1}{k_{F2}}\mathrm{d}\varphi-\int_{-\pi-\alpha}^{\alpha}\frac{1}{k_{F1}}\mathrm{d}\varphi\right) \\
  &=-\frac{\alpha}{\pi}\frac{\sin3\theta}{|\sin3\theta|}-\frac{\delta}{4\pi}\frac{\mu_B^3g^3B^3\sin3\theta}{8\lambda^4}\int_{0}^{2\pi}\frac{1}{k_{F2}}\mathrm{d}\varphi \\
  &=-\frac{\alpha}{\pi}\frac{\sin3\theta}{|\sin3\theta|} \ .
\end{align}
But, we find that the Fermi surfaces given in Eq(\ref{eq:FShard1}) and Eq(\ref{eq:FShard2}) neglect the region between dashed line and solid line as shown in inset of Fig.\ref{fig:fs}(c) where the first term of Eq(\ref{eq:integratek}) is extremely large. Therefore, we have to calculate it additionally. The Fermi surface of this region is approximated by 
\begin{align}
  k^*_{F1}&\approx\left\{\begin{array}{ll}\frac{\frac{\delta \mu_B^3g^3B^3|\sin 3\theta|}{8\lambda^4}}{\sqrt{-1+\gamma^2\sin^2(\varphi-\theta)}}, & \mathrm{if} \ \varphi-\theta\in(-\pi+\alpha,-\alpha)\cup (\alpha,\pi-\alpha) \\
  0, &\mathrm{otherwise}\end{array}\right. ,  \\
  k^*_{F2}&\approx 0 \ .
\end{align}
Plugging them into Eq(\ref{eq:integratek}) gives
\begin{align}
  \sigma_{xy}^{\mathrm{add}}&\approx-\frac{1}{4\pi}\left(\int_{-\pi+\alpha}^{-\alpha}+\int_{\alpha}^{\pi-\alpha}\right)\left[\frac{\frac{\delta\mu_B^3g^3B^3\sin3\theta}{8\lambda^3}}{\sqrt{\lambda^2 k^2 + \frac{\delta^2\mu_B^6g^6B^6\sin^23\theta}{64\lambda^6}}}\right]_{k_{F1}^*}^0 \mathrm{d}\varphi \\
  &\approx\frac{1}{2\pi}\int_{-\pi+\alpha}^{-\alpha}\left(\frac{\frac{\delta\mu_B^3g^3B^3\sin3\theta}{8\lambda^3}}{\sqrt{\lambda^2 {k_{F1}^*}^2 + \frac{\delta^2\mu_B^6g^6B^6\sin^23\theta}{64\lambda^6}}}-\frac{\sin3\theta}{|\sin3\theta|}\right) \mathrm{d}\varphi \\
  &\approx\left(\frac{\alpha}{\pi}-\frac{1}{2}\right)\frac{\sin3\theta}{|\sin3\theta|}+\frac{1}{2\pi}\frac{\sin3\theta}{|\sin3\theta|}\int_{-\pi+\alpha}^{-\alpha}\sqrt{1-\frac{1}{\gamma^2\sin^2\varphi}}\mathrm{d}\varphi \ .
\end{align}
By taking substitution $\gamma^2\sin^2\varphi=1+(\gamma^2-1)\sin \varphi^*$ and transforming it to complex contour integral, the integral can be derived. 
\begin{align}
  \int_{-\pi+\alpha}^{-\alpha}\sqrt{1-\frac{1}{\gamma^2\sin^2\varphi}}\mathrm{d}\varphi=\pi\left(1-\frac{1}{\gamma}\right) .
\end{align}
At last we obtain the additional part as
%$\int_{-\pi+\alpha}^{-\alpha}\sqrt{1-a^2\csc^2\varphi}\mathrm{d}\varphi=2\int_{0}^{\pi/2}\frac{(1-a^2)\sin^2\varphi^*}{a^2+(1-a^2)\sin^2\varphi^*}\mathrm{d}\varphi^*=\pi(1-a)$
\begin{align}
  \sigma_{xy}^{\mathrm{add}} &\approx\left(\frac{\alpha}{\pi}-\frac{1}{2\gamma}\right)\frac{\sin3\theta}{|\sin3\theta|} \ . 
\end{align}
Finally we obtain the Hall conductance as
\begin{align}
  \sigma_{xy}\approx\sigma^{0}_{xy}+\sigma^{\mathrm{add}}_{xy}=-\frac{1}{2\gamma}\frac{\sin3\theta}{|\sin3\theta|}, \ \mathrm{if}\ \mu=E^* \ \mathrm{and} \ \gamma>1\ .
\end{align}
In summary the intrinsic in-plane AHE conductance for this 2D electron gas with $C_3v$ symmetry is 
\begin{align}
  \sigma_{xy}\approx
  \left\{\begin{array}{ll}
    -\frac{ \delta \mu_B^3g^3B^3\sin 3\theta}{16\mu\lambda^3}\left(1 + \frac{\mu_B^2g^2B^2\hbar^2}{8\mu\lambda^2m} + \frac{\mu_B^4g^4B^4\hbar^4}{64\mu^2\lambda^4m^2}\right), & \mathrm{if}\ \mu\gg E^* \\
    -\frac{1}{2}\frac{\sin 3 \theta}{|\sin 3 \theta|}+\frac{\delta\mu_B^2g^2B^2}{16\lambda^3}\frac{\sin3\theta}{\sqrt{1/\gamma^2-1}}, & \mathrm{if}\ \mu=E^* \ \mathrm{and} \ \gamma<1 \\
    -\frac{1}{2\gamma}\frac{\sin3\theta}{|\sin3\theta|}, & \mathrm{if}\ \mu=E^* \ \mathrm{and} \ \gamma>1
  \end{array}\right. \ .
\end{align}
\end{widetext}
\end{document}